# Simulation-based digital twinning of activation and repolarisation sequences from the ECG across healthy and diseased hearts

James A Coleman[a], Julia Camps[b], Abdallah I Hasaballa[a], Alfonso Bueno-Orovio[a,*]

[a]Department of Computer Science, University of Oxford, Oxford, United Kingdom

[b]Department of Engineering, Universitat Pompeu Fabra, Barcelona, Spain

*Corresponding author: Alfonso Bueno-Orovio alfonso.bueno@cs.ox.ac.uk, Department of Computer Science, University of Oxford, Oxford, United Kingdom

# Abstract

Abnormal patterns of ventricular repolarisation are thought to contribute to lethal arrhythmias in various cardiac conditions, including inherited and acquired channelopathies, cardiomyopathies, and ischaemic heart disease. However, methods to detect these repolarisation abnormalities are limited.

In this study, we introduce and assess a novel simulation-based method to infer ventricular activation and repolarisation times from the 12-lead electrocardiogram (ECG) and magnetic resonance-derived ventricular anatomical reconstruction, applicable for the first time to both healthy controls and cases with abnormal repolarisation.

First, ventricular activation times were reconstructed through iterative refinement of early activation sites and conduction velocities, until the model and target QRS complexes matched. Then, ventricular repolarisation times were reconstructed through iterative refinement of ventricular action potential durations and an action potential shape parameter until the model and target T waves matched, including regularisation. Repolarisation inference was evaluated against 18 benchmark simulations with known repolarisation times, including both control and hypertrophic cardiomyopathy (HCM) cases with abnormal repolarisation.

Inferred repolarisation times showed good agreement with the ground truth in control and HCM (Spearman $r = 0.63 \pm 0.11$ and $0.65 \pm 0.19$, respectively), with the inferred model T waves closely matching the target T waves ($r = 0.81 \pm 0.05$ and $0.78 \pm 0.08$, respectively). The method further demonstrated flexibility in reconstructing the macroscopic patterns of delayed repolarisation across a range of abnormal ventricular repolarisation sequences, demonstrating applicability to a wide range of pathological scenarios.

Simulation-based inference can accurately reconstruct repolarisation times from the 12-lead ECG in cases with both normal and abnormal repolarisation patterns.

# 1. Introduction

Lethal arrhythmias may be attributed to pathological patterns of ventricular repolarisation in a range of cardiac diseases, including both inherited and acquired channelopathies (such as congenital or drug-induced long QT syndrome), cardiomyopathies, and ischaemic heart disease (Juárez et al., 2023). It is thought that the assessment of ventricular repolarisation characteristics in such disorders could aid patient risk stratification and identify those amenable to pharmacologic modulation of repolarisation (Coleman et al., 2024a). However, non-invasive electrophysiological assessment methods remain limited, in part due to the complexity, non-uniqueness and instability of the inverse electrocardiography (ECG) problem (Pullan et al., 2010). Even using ECG imaging, where electrophysiological data is derived from tens to hundreds of vest-worn electrodes, repolarisation is still typically only characterised at the epicardial surface, missing the myocardium and interventricular septum. This is particularly limiting in heart diseases with a heterogeneous spatial presentation such as hypertrophic cardiomyopathy (HCM), where the septum manifests most of the disease phenotype and where invasive measurements have highlighted abnormal septal repolarisation patterns as a disease feature (Coppini et al., 2013; Joy et al., 2024).

Accurate reconstruction of such abnormal repolarisation patterns is crucial for the cardiac digital twin vision, which aims to combine patient clinical data with advanced computational modelling, to simulate and predict cardiac function *in silico* (Cluitmans et al., 2024; Corral-Acero et al., 2020; Laubenbacher et al., 2024). As a multiphysics construction, the cardiac digital twin encompasses electrical, mechanical, and structural modelling of the heart, enabling personalised simulation across cardiac functions to support diagnosis, risk stratification, and therapy planning. In the electrophysiological context, this may include *in silico* arrhythmia induction pacing protocols to characterise arrhythmic risk, and *in silico* evaluation of both device-based and pharmacologic therapies (Coleman et al., 2024a; Qian et al., 2021; Roney et al., 2018).

Simulation-based inference of ventricular electrophysiological properties is emerging as a promising tool to inform the cardiac digital twin models that may even outperform ECG imaging (Campos et al., 2024; Sánchez et al., 2025), where clinical ECG data is integrated with prior electrophysiological knowledge to characterise behaviour throughout the ventricles, including the septum. Briefly, such approaches iteratively refine electrophysiological function through tweaking a number of activation and/or repolarisation *in silico* model parameters until a plausible ventricular electrophysiology model is devised, yielding an ECG matching that of the patient (Camps et al., 2025, 2021; Grandits et al., 2025; Pezzuto et al., 2021). This first relies on the patient ventricular anatomy being reconstructed from cardiac magnetic resonance (CMR) images (Banerjee et al., 2021), such as to produce a patient-specific mesh on which the electrophysiological function can then be optimised to fit the patient ECG.

Recent developments have shown that the ventricular activation sequence can be accurately inferred from the QRS complex and CMR-derived ventricular anatomy, where performance of the method was well evaluated against known ground-truth ventricular activation times (Camps et al., 2021; Pezzuto et al., 2021). It has further been shown that it is possible to infer ventricular repolarisation properties that are partially consistent with the patient T wave in healthy cases (Camps et al., 2025; Gillette et al., 2021), but the repolarisation times inferred with such methods have not been thoroughly evaluated against known ground-truth ventricular repolarisation times. Moreover, such methods have previously imposed monotonic spatial gradients in ventricular action potential duration (APD) gradients along the apicobasal, anteroposterior, interventricular, and transmural axes. This means that repolarisation patterns which do not conform to these expected spatial distributions cannot be inferred from the T wave, limiting applicability to cardiac disease. For example, repolarisation patterns confined to the septum in HCM may be inadequately reproduced using a single apicobasal APD gradient parameter (Lyon et al., 2018).

The present study therefore aimed to (1) develop a novel computational inference framework capable of reconstructing ventricular repolarisation patterns in both healthy *and* spatially heterogeneous disease cases (such as HCM); and (2) evaluate the accuracy of inferred repolarisation times by comparing them to ground-truth repolarisation times derived from biophysically detailed cardiac electrophysiology benchmark simulations. This notably extends upon previous works by developing methods to infer repolarisation patterns where it is most clinically needed –in cardiac disease where abnormal repolarisation is present– and by comparing inferred repolarisation models against the ground truth source of the ECG signals, a crucial empirical benchmark sometimes overlooked (Nash and Pullan, 2005).

# 2. Methods

This section first details the generation of benchmark ventricular electrophysiology simulations (Section 2.1), where the derived simulated benchmark ECGs were then used as target ECGs in the inference method (later detailed in Section 2.2).

## 2.1 Benchmark human biventricular electrophysiology simulations

The capability of the inference method to infer activation and repolarisation properties from the 12-lead ECG was evaluated on a set of 18 benchmark human biventricular cardiac electrophysiology simulations, performed using six hexahedral meshes derived from four control and two HCM subjects (Supplementary Figure S1) (Lyon et al., 2018). This represents a significant expansion in the evaluation of simulation-based inference methods, where previous works considered only a smaller number of 1-3 benchmark simulations derived from only healthy subjects (Camps et al., 2025; Grandits et al., 2025; T. Grandits et al., 2024).

Benchmark monodomain simulations were performed using the MonoAlg3D cardiac electrophysiology simulator (Sachetto Oliveira et al., 2018) coupled to the ToR-ORd human ventricular action potential (AP) model (Tomek et al., 2019). The ToR-ORd model uses biophysically detailed ion channel behaviours to produce simulated APs and calcium transients, and has previously been extensively validated to reproduce human AP characteristics under both healthy and diseased conditions, including delayed repolarisation in HCM. MonoAlg3D facilitates the integration of this cellular AP model at the ventricular scale with diffusion, from which spatial activation and repolarisation patterns can be derived alongside realistic ECGs (Coleman et al., 2024b). In the absence of invasive clinical measurements, the benchmark simulations offer a credible source of ground truth against which the inference method accuracy can be assessed.

All benchmark simulations incorporated transmural heterogeneity (70% endo-, 30% epicardial cells), apicobasal APD gradients, and human conduction velocities within $[30,70]$ $cm/s$, similar to previous work (Coleman et al., 2024a). Ventricular activation sequences in the six meshes were defined similar to as in previous work (Cardone-Noott et al., 2016), where varying root nodes were defined on the endocardium, then each endocardial surface node was assigned a stimulus time proportional to its distance to its closest root node, as computed with Dijkstra's algorithm. All resulting activation sequences were consistent with the findings of Durrer *et al.* (Durrer et al., 1970). Benchmark simulations are further detailed in Supplementary Section 1.1.

Each of the six unique biventricular meshes was assigned two repolarisation variants (Supplementary Figure S1): control (with healthy electrophysiology consisting of epi-endocardial and apicobasal APD gradients) and HCM (all the aforementioned APD gradients *and* an HCM-remodelled septal region with prolonged APD secondary to ionic remodelling). Ionic remodelling in HCM affects multiple channels, with the main effects on APD arising due to up-regulation of late $Na^+$ channels and down-regulation of $K^+$ rectifier channels (Coppini et al., 2013). The HCM variant benchmarks had ionic remodelling applied as a sphere of radius $2\ cm$ subjected to a maximal degree of ionic remodelling, encircled by a border of radius $3\ cm$ containing a linear gradient in rescaled ionic conductances from non-remodelled up to maximally remodelled. This was applied centred on the LV endocardial mid-septum, as in previous work (Coleman et al., 2024a, 2024b). The maximal degree of ionic remodelling was set such as to produce simulated maximum $JT_c$ intervals in line with clinical maximum $JT_c$ intervals in HCM, as measured previously (Coleman et al., 2024a).

After pacing single-cell AP simulations to a steady state, the benchmark simulations were paced for a further two beats at 1 Hz. Ground truth activation times and repolarisation times following the second beat were computed for each benchmark simulation, to later be used to evaluate the accuracy of activation and repolarisation times inferred from each benchmark simulation ECG.

For each 12-lead ECG derived from the benchmark simulations, the subset of the ECG used as the target ECG was manually selected for both the activation inference (from the onset of the Q wave to the end of the S wave, in the lead with the longest and most discernible QRS complex) and for the repolarisation inference (from the end of the S wave to the end of the T wave, as in the lead with the longest and most discernible T wave).

## 2.2 Inference of activation and repolarisation sequences from the ECG

### 2.2.1 Overview

The inference method used is a novel approach that combines rapid forward ECG simulations with a discrepancy metric to compare with the target ECG, to iteratively refine estimates of ventricular activation and repolarisation times. The key novelties of this approach are that (1) no global ventricular APD gradients are enforced, such that both healthy and abnormal repolarisation patterns can be inferred (see later Section 2.2.3); and (2) a novel rapid repolarisation surrogate is used to facilitate rapid forward T wave simulations (see later Section 2.2.4). Finally, the inference results are empirically evaluated against known ground truths.

The inference of ventricular electrophysiological properties from the ECG was separated into two broad steps: (1) inference of the activation sequence from the target QRS complex (Figure 1A), followed by (2) inference of the repolarisation sequence from the target T wave (Figure 1B). Both steps used the same iterative method to refine parameters related first to activation, then subsequently parameters related to repolarisation. The selected target QRS complex and T wave subsets from the benchmark electrophysiology simulations from Section 2.1 were used as inputs to the activation and repolarisation inferences, respectively.

The inference method generates pseudo-ECGs corresponding to activation/repolarisation parameters (see later Sections 2.2.2 and 2.2.3), and continuously refines those parameters to better fit the target ECG. Briefly, the inference steps are:

1. Prepare $N$ parameter sets
2. Compute pseudo-ECGs corresponding to these parameter sets
3. Rank generated pseudo-ECGs by discrepancy score compared to the target ECG
4. Replace the parameter sets corresponding to the $p\%$ worst match to the target ECG with mutated copies of parameter sets corresponding to the $(1-p)\%$ best match to the target ECG. Repeat from step 2 until the stopping condition is met.

The process was stopped when the per-iteration decrease in discrepancy score between the inferred and target ECG fell below a set threshold, quantified using a moving-average per-iteration decrease of the median ECG discrepancy score among the $N$ parameter sets. Consequently, the method discards activation/repolarisation parameter sets with poor discrepancy scores between the inferred and target ECG, focusing on resampling solutions with a better ECG discrepancy score. By continuously refining the population of $N$ parameter sets at each iteration, rather than just a single model, the method promotes finding a global minimum rather than prematurely focusing on a single local minimum solution. Further details are available in Supplementary Section 1.2.

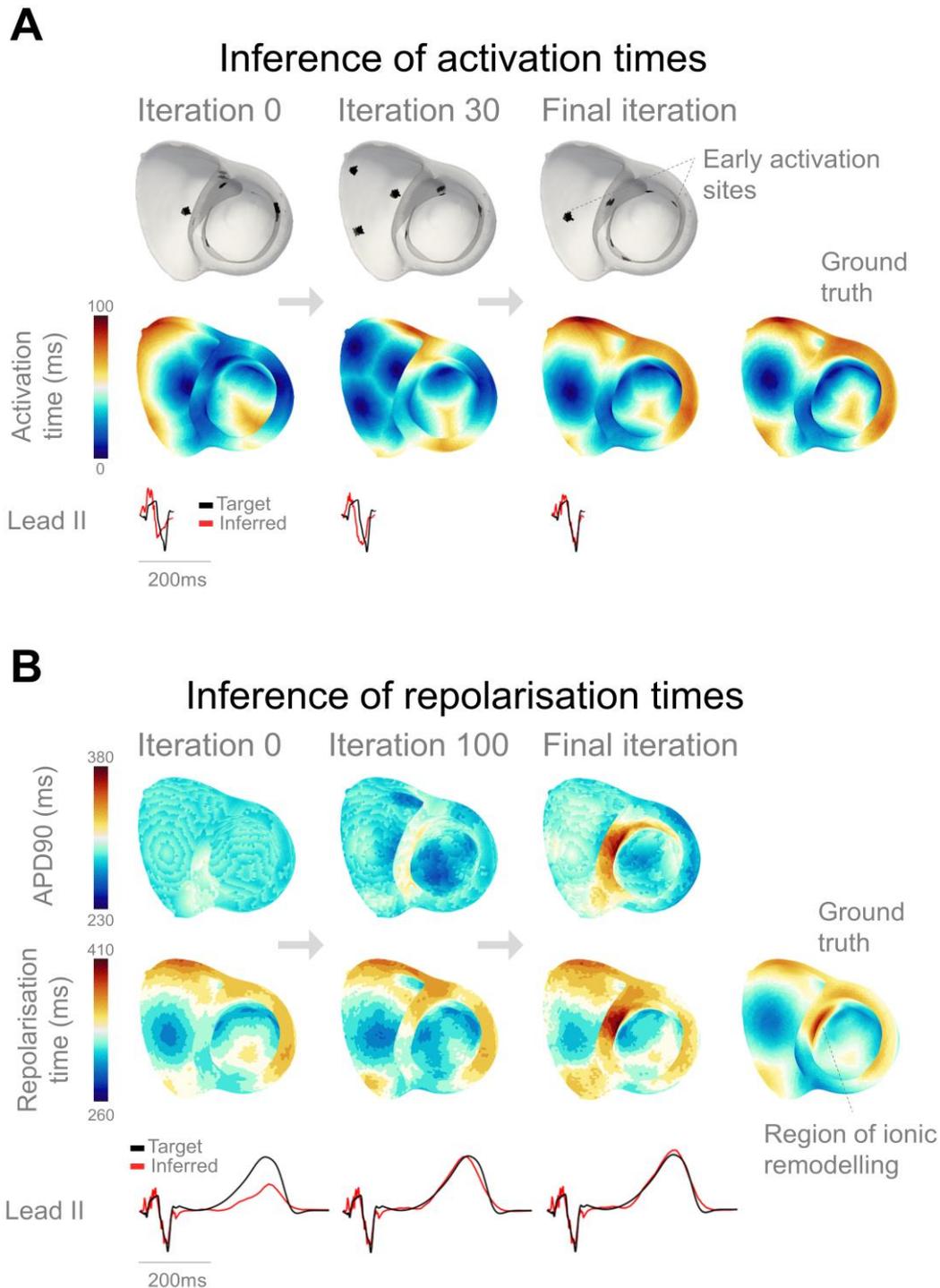

**Figure 1. Iterative refinement of model ventricular activation and repolarisation sequences to match the target 12-lead ECG.** (A) Progression of early activation sites, activation times, and QRS complex of the best activation model across iterations, as compared to the ground-truth activation times and target QRS complex. (B) Progression of ventricular APDs at 90% repolarisation ($APD_{90}$), repolarisation times, and ST complex of the best repolarisation model across iterations, as compared to the ground-truth repolarisation times and target ECG.

## 2.2.2 Activation

Ventricular activation was parameterised by the endocardial conduction velocity ($v_{endo}$) in [80, 140] $cm/s$, myocardial conduction velocity ($v_{myo}$) in [20,90] $cm/s$, and a set of 6-10 early ventricular activation sites. Early activation sites were uniformly sampled from candidate early activation sites (see below), with $v_{myo}$ and $v_{endo}$ also uniformly randomly chosen from their possible ranges. Candidate activation models were simulated using an Eikonal approach and candidate pseudo-ECGs derived between time zero up until the time of the end of the target QRS complex, as previously selected from the target ECG, in 2 $ms$ time increments.

Similarity of the candidate to the target QRS complexes was quantified using a QRS discrepancy score $D_{QRS}$, which measures the mean absolute difference between normalised candidate and target QRS complexes across all leads and time points. This is given by:

$$D_{QRS} = \frac{1}{N_l N_t^{QRS}} \sum_{l=1}^{N_l} \sum_{t=1}^{N_t^{QRS}} \left| \frac{C_{l,t}^{QRS}}{A_l^{QRS,C}} - \frac{G_{l,t}^{QRS}}{A_l^{QRS,G}} \right|$$

where $N_l$ is the number of leads, $N_t^{QRS}$ is the number of samples per QRS complex, $l = 1, \ldots, N_l$ indexes ECG leads, $t = 1, \ldots, N_t^{QRS}$ indexes time points in the QRS complex, $C_{l,t}^{QRS}$ and $G_{l,t}^{QRS}$ are ECG signals at lead $l$ and time point $t$ for the candidate and target QRS complexes, respectively, and $A_l^{QRS,C}$ and $A_l^{QRS,G}$ are lead $l$ QRS amplitudes for the candidate and target signals, respectively.

During mutation of activation parameters, root node positions and/or conduction velocities were changed as detailed in Supplementary Section 1.3. After reaching the stopping condition, the activation times of the parameter set from the population of size $N$ with the closest QRS to the target QRS were used as the inferred activation sequence.

To allow precomputation of ventricular activation times (and thus rapid simulation of the QRS), candidate early ventricular activation sites were defined by Poisson disk sampling points on the endocardium, set to be a minimum distance of 0.5 $cm$ apart. Possible conduction velocities $v_{myo}$ and $v_{endo}$ were discretised in 10 $cm/s$ increments to also facilitate precomputation of ventricular activation times. Rapid pseudo-ECG simulation of candidate activation sequences was further enabled by approximating AP upstrokes with a Heaviside function, as in previous work (Camps et al., 2021).

## 2.2.3 Repolarisation

The repolarisation inference process used the inferred activation times from the previous step as an input.

Ventricular repolarisation was parameterised by assigning cellwise APD at 90% repolarisation (APD$_{90}$), and a global AP shape parameter in the range [0,1] that controlled AP shape via APD at 50% repolarisation (APD$_{50}$). As shown later in Figure 2A, an AP shape parameter of 0 corresponded to a more triangular AP shape (by selecting the minimum possible APD$_{50}$ for each APD$_{90}$), whereas an AP shape parameter of 1 corresponded to a more rounded and long plateau AP shape (by selecting the maximum possible APD$_{50}$ for each APD$_{90}$). Repolarisation parameter sets are initialised with a randomly chosen APD$_{90}$ in [200,400] $ms$ uniformly applied in the ventricles (with small perturbations to the APD$_{90}$ field to ensure unique initial parameter sets), with an AP shape parameter of 0.5. Candidate repolarisation models are then simulated and candidate pseudo-ECGs derived between the time of the end of the target QRS complex until the time of the end of the target T wave (plus a $50\ ms$ safety factor), as previously selected from the target ECG, simulated in $10\ ms$ time increments.

Similarity of the candidate and target T waves was quantified using a T wave discrepancy score $D_T$, which measures the mean absolute difference been candidate and target T waves across all leads and time points, after rescaling the candidate ECG to match the target QRS amplitude and scaling by target T wave amplitude. A regularisation term was further added to each T wave discrepancy score, to enforce spatially smooth solutions (Supplementary Section 1.6). The T wave discrepancy score $D_T$ is given by:

$$D_T = \frac{1}{N_l N_t^T} \sum_{l=1}^{N_l} \sum_{t=1}^{N_t^T} \frac{1}{A_l^{T,G}} \left| \frac{A_l^{QRS,G}}{A_l^{QRS,C}} C_{l,t}^T - G_{l,t}^T \right| + \lambda \langle |\nabla V_m| \rangle_{x,t}$$

where $\lambda$ is the regularisation parameter, $\langle |\nabla V_m| \rangle_{x,t}$ is the mean absolute gradient of membrane potential of the candidate repolarisation model averaged over space and time, $N_l$ is the number of leads, $N_t^T$ is the number of samples per T wave, $l = 1, \ldots, N_l$ indexes ECG leads, $t = 1, \ldots, N_t^T$ indexes time points in the T wave, $A_l^{T,G}$ is lead $l$ T wave amplitude for the target signal, $C_{l,t}^T$ and $G_{l,t}^T$ are ECG signals at time point $t$ and lead $l$ for the candidate and target T waves, respectively, and $A_l^{QRS,C}$ and $A_l^{QRS,G}$ are lead $l$ QRS amplitudes for the candidate and target signals, respectively.

During mutation of repolarisation parameters, ventricular $APD_{90}$s and the global AP shape parameter were changed as detailed in Supplementary Section 1.4. Importantly, the mutation step involved selecting a random ventricular mesh site, then changing local $APD_{90}$s within some Dijkstra distance of the site by a given multiplier. Over many iterations, the repolarisation time field is progressively modified to produce a T wave matching that of the target ECG, where spatially heterogeneous repolarisation patterns can be reconstructed. This means that inferred $APD_{90}$ gradients were not imposed across any ventricular axis, but emerged during the mutation step and were retained subject to the ECG discrepancy score. After reaching the stopping condition, the repolarisation times of the parameter set from the population of size $N$ with the closest match to the target T wave (and accounting for the regularisation term) were used as the inferred repolarisation sequence.

### 2.2.4 Fast repolarisation surrogate model

The iterative refinement of ventricular repolarisation properties requires the simulation of many candidate repolarisation models and corresponding ECGs, which would be computationally infeasible to perform using a biophysically detailed electrophysiological model such as that used for the benchmark simulations in Section 2.1. Therefore, an optimised fast surrogate model is proposed and used in candidate models during the repolarisation inference.

Rapid pseudo-ECG simulation of candidate repolarisation sequences was facilitated by approximating the reaction-diffusion behaviour of the monodomain equation with a precomputed AP table (Fig. 2A) and Gaussian smoothing of membrane potentials, expanding previous work (Camps et al., 2025). The AP table, indexed by $(APD_{90}, APD_{50})$, comprised 6,606 unique AP morphologies precomputed using the ToR-ORd ventricular cardiomyocyte electrophysiology model (Tomek et al., 2019). APs contained in the table had $APD_{90}$s in the range $[200,400]\ ms$ in $1\ ms$ increments, and a viable range of $APD_{50}$s for each given $APD_{90}$, consistent with cellular ToR-ORd model dynamics. For example, the AP table row corresponding to the minimum $APD_{90}$ of $200\ ms$ had $APD_{50}$s in $[155,169]\ ms$, while the maximum APD$_{90}$ row of $400\ ms$ had APD$_{50}$s in $[292,341]\ ms$, in $1\ ms$ increments. The AP table was constructed by varying the maximal conductances of I$_{Kr}$ and I$_{K1}$ between $40-250\%$ of their default values in the ToR-ORd model. Compared to previous work in which only I$_{Ks}$ conductances were varied to precompute an AP table solely indexed by APD$_{90}$ (Camps et al., 2025), the novel use of I$_{Kr}$ and I$_{K1}$ in the present study enabled computation of AP shapes with variable plateau shape for each APD$_{90}$, facilitating an AP table indexed by $(APD_{90}, APD_{50})$.

For each pseudo-ECG simulation of candidate repolarisation sequences, membrane potentials from the AP table were first placed into the ventricular mesh corresponding to the local APD$_{90}$ and APD$_{50}$ parameters. To approximate diffusion, Gaussian smoothing was then applied to allocated membrane potentials, with a correction factor to account for the complex ventricular geometry. The smoothed membrane potential at mesh site $i$ and time $t$, denoted as $\tilde{V}_i(t)$, is given by:

$$\tilde{V}_i(t) = \frac{\sum_{j \in \Omega} G_\sigma(i-j) V_j(t)}{\sum_{j \in \Omega} G_\sigma(i-j)}$$

where $\Omega$ is the spatial domain of the ventricular mesh, $V_j(t)$ is the pre-smoothing membrane potential at mesh site $j$ and time $t$, and $G_\sigma(i-j)$ is the value of the Gaussian kernel calculated as:

$$G_\sigma(i-j) = \exp\left(-\frac{\|\vec{r}_i - \vec{r}_j\|^2}{2\sigma^2}\right)$$

where $\|\vec{r}_i - \vec{r}_j\|$ is the Euclidean distance between points $i$ and $j$, and $\sigma$ is the smoothing scale parameter fixed over space and time. The smoothing scale parameter for each repolarisation inference was determined by the inferred myocardial conduction velocity $v_{myo}$, where the relationship between $v_{myo}$, myocardial conductivity, and optimal smoothing scale parameter was previously measured in MonoAlg3D for cable and slab simulations with conductivities in $[0,6]\ mS/cm$ (Fig. 2B). The relationship between $v_{myo}$ and the optimal smoothing scale parameter is summarised in Figure 2C. The initial setting of ventricular APD$_{90}$s from the AP table, combined with Gaussian smoothing of membrane potentials in the repolarisation phase, meant that the true APD$_{90}$ distribution (from which pseudo-ECGs were derived) was effectively smoothed (Fig. 2D).

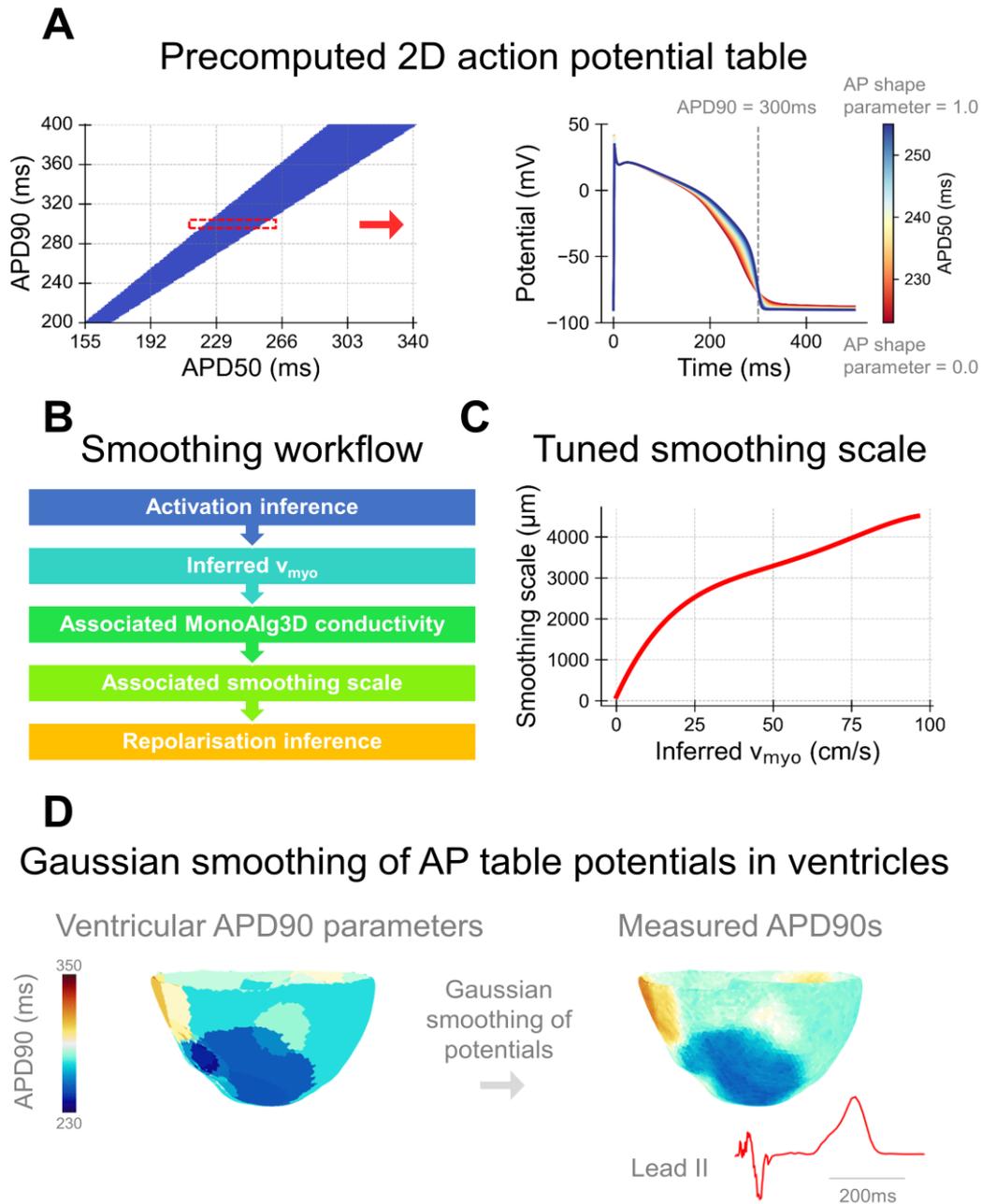

**Figure 2. Methods used to facilitate rapid pseudo-ECG simulation of candidate repolarisation models.** (A) Possible ($APD_{90}$, $APD_{50}$) combinations used to index AP shapes in the bivariate AP table (left), shown alongside an example row of APs from the table with ($APD_{90} = 300\,ms$, $APD_{50}$) (right) illustrating that the AP table contains a range of AP shapes for each $APD_{90}$. (B) Workflow for the derivation, from the activation inference, of the smoothing scale applied to membrane potentials during repolarisation. (C) Relationship between inferred $v_{myo}$ and the smoothing scale applied during repolarisation. (D) Illustration of how cellwise $APD_{90}$s in the ventricles (left) change post-smoothing of membrane potentials from the AP table (right), where ECGs were derived from the post-smoothing repolarisation model.

## 2.3 Statistical methods

Data are expressed as mean ± standard deviation. Similarity between inferred and ground truth activation/repolarisation times was quantified using the mean absolute difference across ventricular cells and Spearman correlation coefficient.

# 3. Results

In total, 76 inference runs were completed. This includes 6 activation inferences (Section 3.1), 12 repolarisation inferences split evenly across control and septal HCM cases (Section 3.2), 6 repolarisation inferences for 6 additional varied HCM cases (Section 3.2), 12 further repolarisation inferences across control and septal HCM to assess propagation of activation inference error to repolarisation (Section 3.3), and 40 repolarisation inferences used to tune the regularisation parameter (Supplementary Section 1.6). Each inference run constituted approximately 10 hours runtime, distributed across 12 cores using the University of Oxford Advanced Research Computing facility. Each inference run typically took approximately 600 iterations to reach the stopping condition, corresponding to approximately 80,000 candidate models per inference.

## 3.1 Activation times inferred from the QRS complex

Results of the inference process applied to ventricular activation are shown in Figure 3. The inferred and ground truth spatial distributions of ventricular activation times were highly similar (Fig. 3A), with mean absolute errors of $6.5 \pm 1.7\ ms$ and averaged Spearman correlation coefficients of $0.84 \pm 0.10$ for the six benchmarks (Fig. 3B). As shown in Figure 3C, mean absolute errors were greater in the right ventricle (RV) than in left ventricle (LV) ($7.9 \pm 1.9$ vs. $5.9 \pm 1.8\ ms$, respectively). The inferred activation models produced QRS complexes in close agreement with the target QRS complexes, with Spearman correlation coefficients of $0.95 \pm 0.02$ (Fig. 3D). This constituted important verification of previous works in which activation times were accurately inferred from the QRS complex (Camps et al., 2021), facilitating the extension to inferring repolarisation times from the T wave. Full results of the activation times inferences across the six benchmarks are shown in Supplementary Figure S4.

**Figure 3. Ventricular activation times inferred from the QRS complex.** (A) Inferred activation times (top) vs. ground truth activation times (bottom) for a representative benchmark, shown from anterior, superior and posterior viewpoints. (B) Mean absolute errors and Spearman correlation coefficients between inferred and ground truth activation times for the six benchmarks. (C) Mean absolute errors compared between LV and RV. (D) Representative target and inferred lead II QRS complexes and mean Spearman correlation coefficients derived from the 12 leads.

## 3.2 Repolarisation times inferred from the T wave

Results of the inference process applied to ventricular repolarisation in control cases are shown in Figure 4. The inferred and ground truth spatial distributions of ventricular repolarisation times were similar (Fig. 4A), with mean absolute errors of $11.0 \pm 1.2\ ms$ and averaged Spearman correlation coefficients of $0.63 \pm 0.11$ for the six control benchmarks (Fig. 4B). As shown in Figure 4C, mean absolute errors between the RV and LV were variable between benchmarks and similar overall ($11.4 \pm 1.4$ vs. $10.9 \pm 1.8\ ms$, respectively). The inferred repolarisation models produced T waves in close agreement with the target T waves, with Spearman correlation coefficients of $0.81 \pm 0.05$ (Fig. 4D). Full results of the repolarisation times inferences across the six benchmarks with control electrophysiology are shown in Supplementary Figure S5. A representative 12-lead ECG of an inferred control model as compared to its target 12-lead ECG is further shown in Supplementary Figure S7.

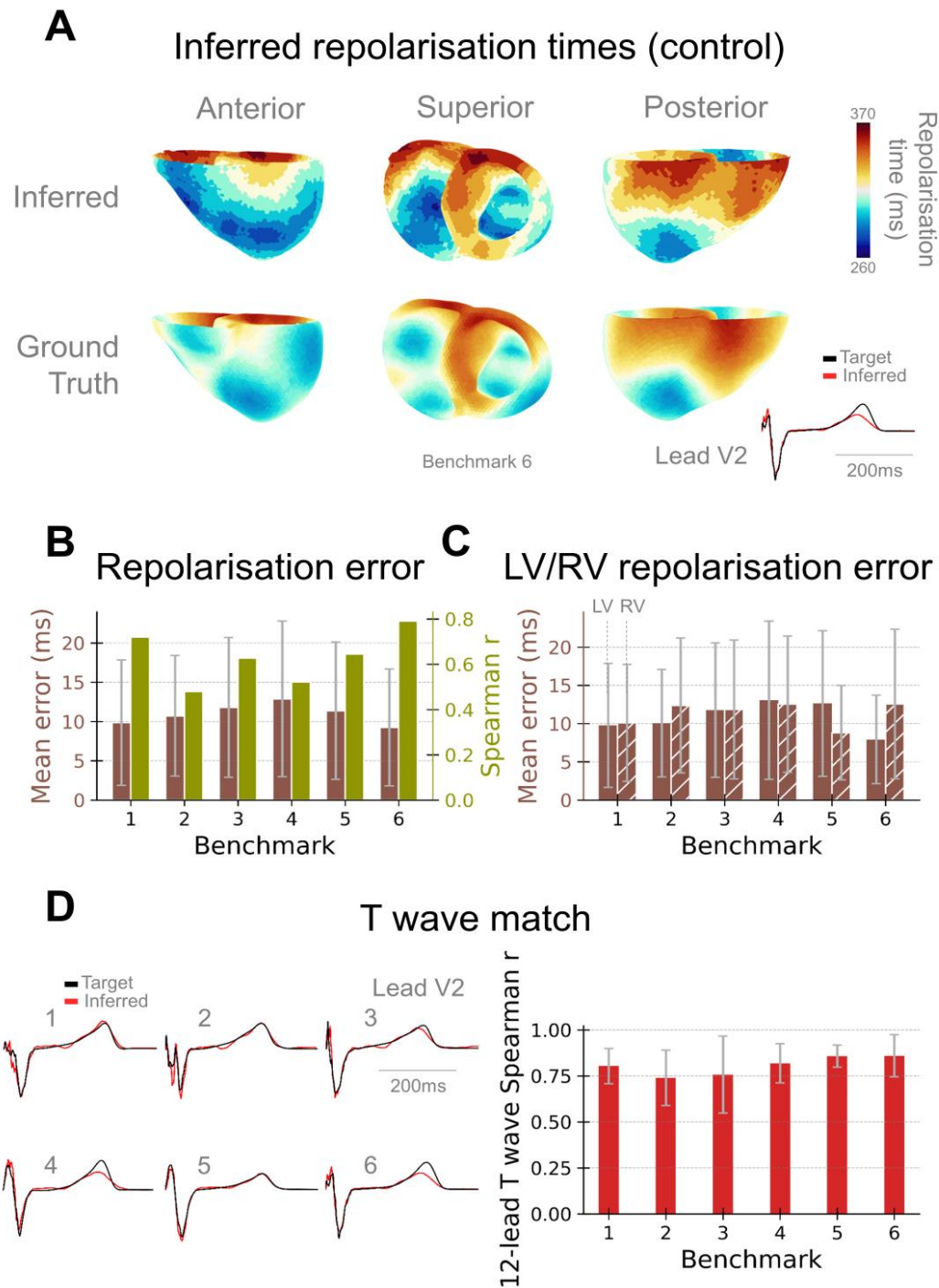

**Figure 4. Ventricular repolarisation times inferred from the T wave in control cases.** (A) Inferred repolarisation times (top) vs. ground truth repolarisation times (bottom) for a representative control benchmark, shown from anterior, superior and posterior viewpoints. (B) Mean absolute errors and Spearman correlation coefficients between inferred and ground truth repolarisation times for the six control benchmarks. (C) Mean absolute errors compared between LV and RV. (D) Representative target and inferred lead V2 ECGs and mean Spearman correlation coefficients (comparing T waves only) across the 12 leads.

Of note, under healthy (control) conditions, the spatial distribution of repolarisation times resembled the spatial distribution of activation times (Fig. 4A vs. Fig. 3A) for both the inferred and ground truth repolarisation times, as reported to occur experimentally (Myles et al., 2010). This is explained by the timescales over which ventricular activation occurs, relative to the typical timescales of APD gradients such as epi-endocardial differences. Although repolarisation times are the summation of both local activation times and $APD_{90}$s, ventricular activation occurring on the order of $100\ ms$ had a stronger effect on repolarisation times than $APD_{90}$ gradients on the order of $20\ ms$. It is only when greater $APD_{90}$ gradients were introduced, such as in the HCM cases on the order of $100\ ms$, that the repolarisation time field deviated more significantly from the activation time field (Fig. 5A vs. Fig. 3A).

Results of the inference process applied to ventricular repolarisation in HCM cases are shown in Figure 5. The inferred and ground truth spatial distributions of ventricular repolarisation times were similar (Fig. 5A), with mean absolute errors of $14.4 \pm 2.5\ ms$ and averaged Spearman correlation coefficients of $0.65 \pm 0.19$ for the six HCM benchmarks (Fig. 5B). As shown in Figure 5C, mean absolute errors between the RV and LV were variable between benchmarks and similar overall ($15.0 \pm 3.4$ vs. $14.2 \pm 3.0\ ms$, respectively). Even in the case of pathological repolarisation, the inferred repolarisation models produced T waves in close agreement with the target T waves, with Spearman correlation coefficients of $0.78 \pm 0.08$ (Fig. 5D). Full results of the repolarisation times inferences across the six HCM benchmarks are shown in Supplementary Figure S6. A representative 12-lead ECG of an inferred HCM model as compared to its target 12-lead ECG is further shown in Supplementary Figure S8.

A key evaluation target of the repolarisation inference in HCM cases was that the method should successfully locate the region of septal ionic remodelling. This was achieved in 5 of the 6 HCM cases, where inferred repolarisation times showed a sizeable septal region with prolonged repolarisation. In the HCM case where prolonged repolarisation was not completely identified in the septum (benchmark 2), prolonged repolarisation was instead attributed to the RV (Supplementary Fig. S10), explaining the lower repolarisation time Spearman correlation coefficient in benchmark 2 (Fig. 5B). This was also reflected by a poorer T wave match (Fig. 5D).

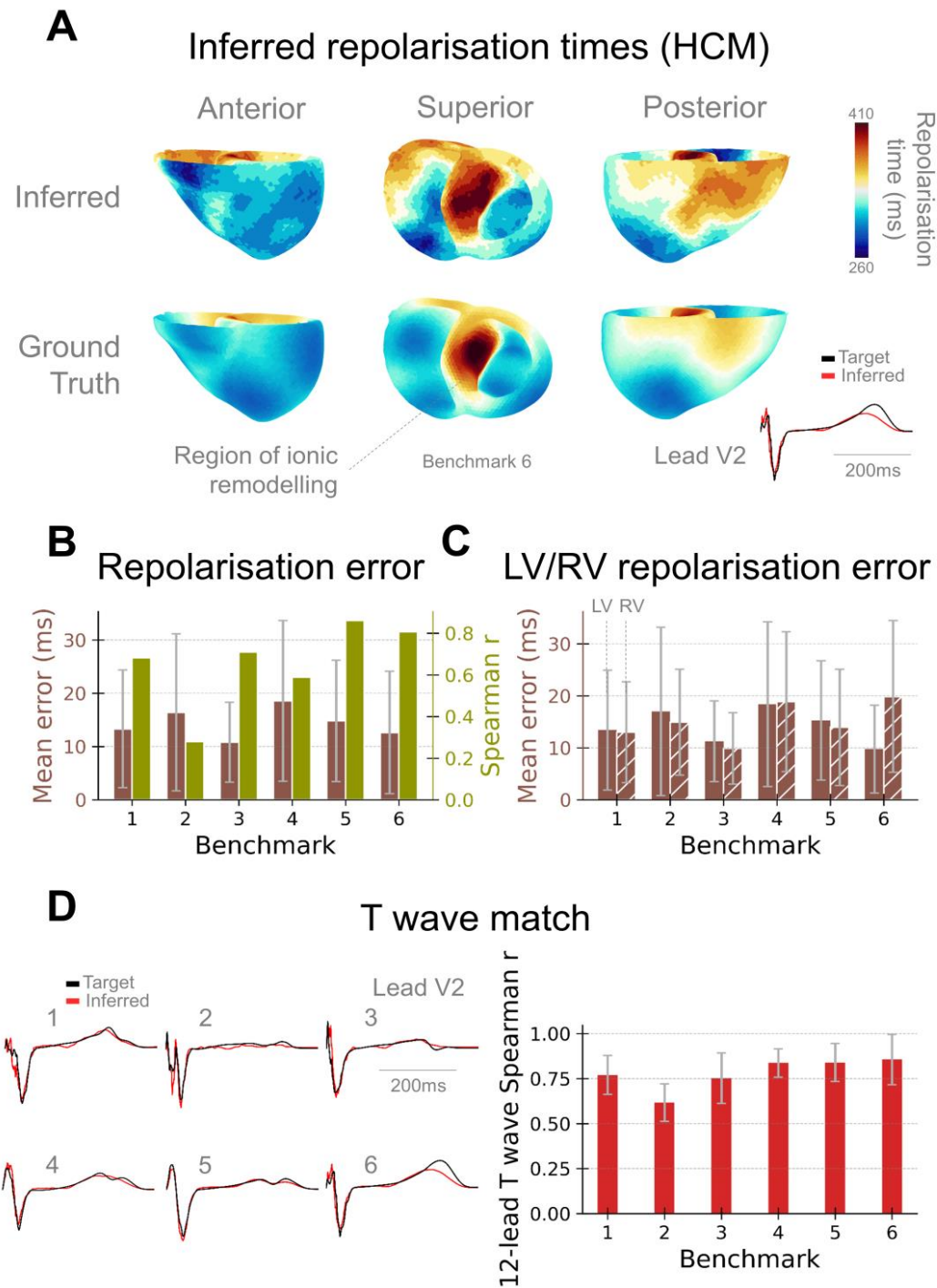

**Figure 5. Ventricular repolarisation times inferred from the T wave in HCM cases.** (A) Inferred repolarisation times (top) vs. ground truth repolarisation times (bottom) for a representative HCM benchmark, shown from anterior, superior and posterior viewpoints. (B) Mean absolute errors and Spearman correlation coefficients between inferred and ground truth repolarisation times for the six HCM benchmarks. (C) Mean absolute errors compared between LV and RV. (D) Representative target and inferred lead V2 ECGs and mean Spearman correlation coefficients (comparing T waves only) across the 12 leads.

The repolarisation inference was evaluated for a further six HCM cases with various affected regions, distinct from the previous six HCM cases in which ionic remodelling was applied as a septal region. These further cases were considered because, although delayed repolarisation is hypothesised to be associated with (typically septal) hypertrophy in HCM, its true spatial distribution is unknown, and hypertrophy can be present in other parts of the ventricles (e.g., in apical HCM). These six additional HCM cases showed good agreement between the inferred and ground truth spatial distributions of pathological ventricular repolarisation times (Fig. 6), with averaged Spearman correlation coefficients of $0.72 \pm 0.08$. Of note, inference was successful even for a non-spherical region of septal ionic remodelling ($r = 0.75$), and when multiple non-spherical regions were affected ($r = 0.70$). Inference was highly successful when the entire ventricular base was affected ($r = 0.83$), but was poorer when the entire LV was affected ($r = 0.56$). Despite prolonged repolarisation being inferred throughout the LV in this case, prolonged repolarisation was inferred also in much of the RV. Finally, spherical regions of ionic remodelling in non-septal regions such as the anterior LV and posterior LV were also reasonably well inferred ($r = 0.70$ and $r = 0.76$, respectively).

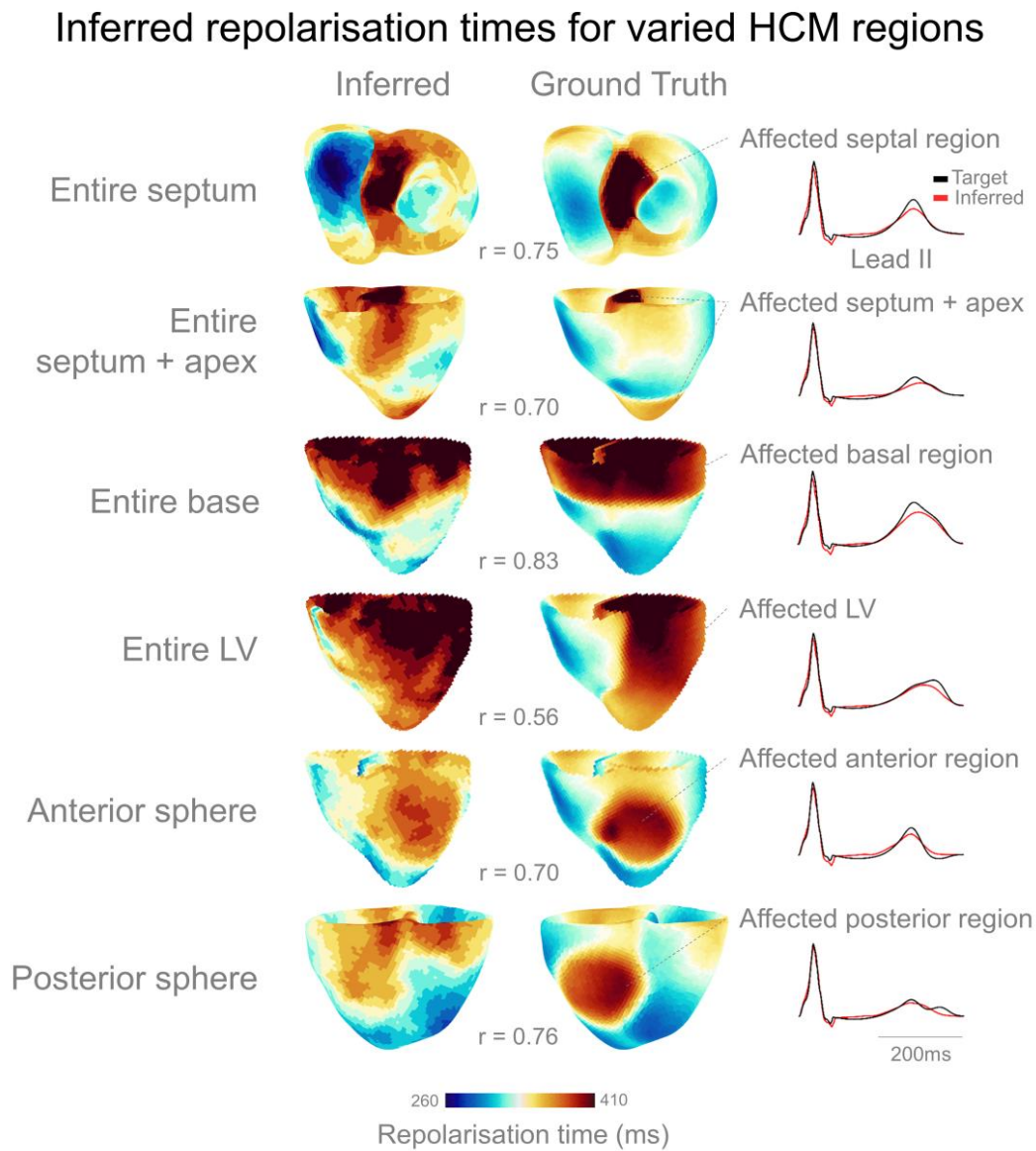

**Figure 6. Ventricular repolarisation times inferred from the T wave for miscellaneous HCM cases with different regions affected by ionic remodelling.** Inferred repolarisation times (left) vs. ground truth repolarisation times (right), alongside representative target and inferred lead II ECGs, for miscellaneous HCM cases with ionic remodelling. Regions affected by ionic remodelling, from top to bottom, are (1) the entire septum (not just a spherical subset as before), (2) the entire septum and the apex, (3) the ventricular base, (4) the entire LV, (5) a spherical subset of the anterior LV, and (6) a spherical subset of the posterior LV. Spearman correlation coefficients between inferred and ground truth repolarisation times are denoted for each case.

## 3.3 Effect of error in inferred activation times on the accuracy of inferred repolarisation times

The six control and six HCM cases used to evaluate the repolarisation inference method previously in Figures 4 & 5 used the activation times inferred in Figure 3 as an input to the repolarisation inference process, because in practice the ground truth activation times would be unknown. However, to evaluate how error in the inferred activation times may propagate to error in inferred repolarisation times, the repolarisation inference was repeated for all the aforementioned cases, but instead using the ground truth activation times as an input.

The similarity between inferred and ground truth repolarisation times, when using the inferred vs. ground truth activation times as an input to the repolarisation inference, was compared as shown in Figure 7. When using the ground truth activation times rather than inferred activation times, repolarisation inference was more accurate with an increase in Spearman correlation coefficients from $r = 0.63 \pm 0.11$ to $r = 0.73 \pm 0.04$ among the six control cases (Fig. 7A, left), and an increase from $r = 0.65 \pm 0.19$ to $r = 0.74 \pm 0.08$ among the six HCM cases (Fig. 7A, right). Therefore, error in the inferred activation times contributed to error in the inferred repolarisation times, variably so among benchmarks. This is illustrated in Figure 7B, where agreement between inferred and ground truth repolarisation times is visibly better in the inferences where ground truth activation times were used, being particularly noticeable in the RV. Interestingly, the HCM case in which the HCM-remodelled region was misidentified as being in the RV had a more accurate repolarisation inference when using the ground truth activation times ($r = 0.63$ vs. $r = 0.28$) (Supplementary Fig. S10).

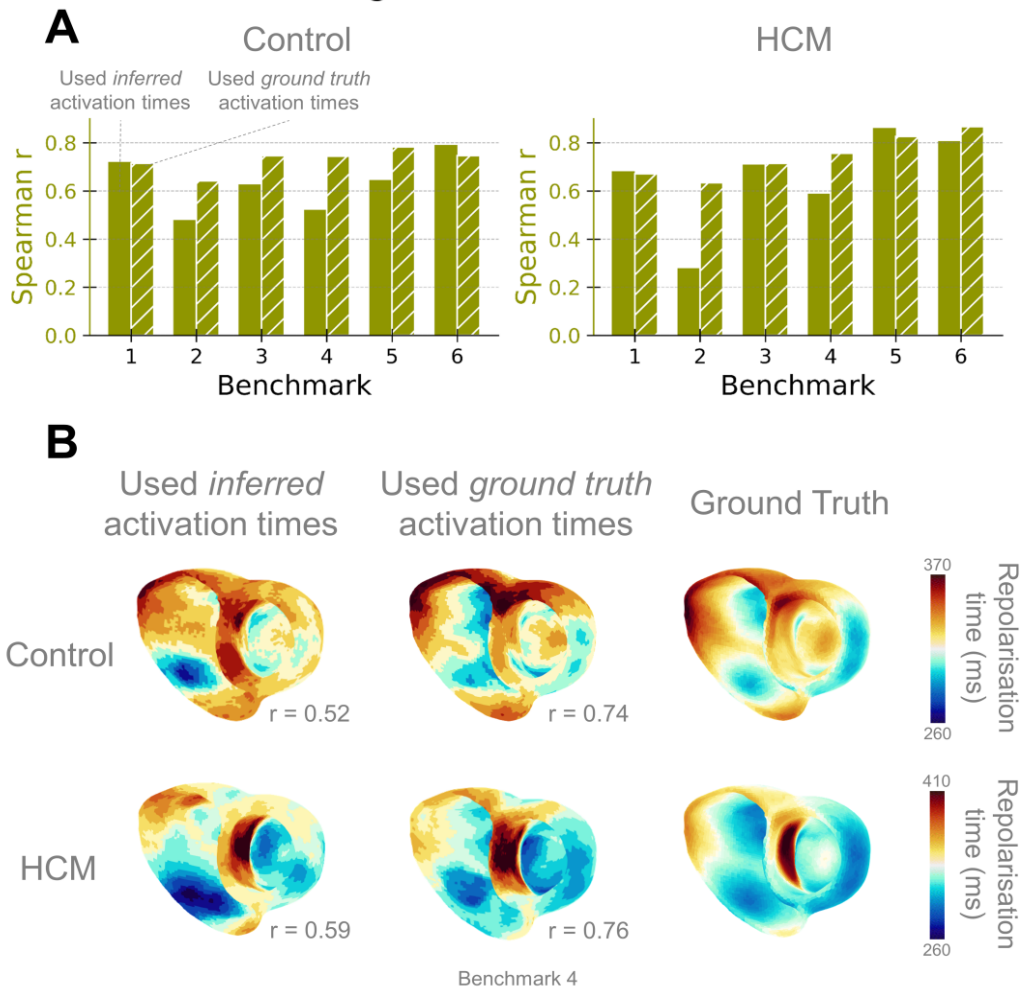

**Figure 7. Ventricular repolarisation times inferred using inferred activation times as an input, vs. those inferred using ground truth activation times as an input.** (A) Spearman correlation coefficients between inferred and ground truth repolarisation times for the six control and six HCM benchmarks, where repolarisation inferences used inferred activation times (solid colour) or ground truth activation times (hatched) as input, for control (left) and HCM (right) cases. (B) Representative effect of using inferred vs. ground truth activation times on inferred ventricular repolarisation times in a control (top) and HCM (bottom) case.

# 4. Discussion

This study aimed to develop and evaluate a simulation-based inference framework capable of reconstructing healthy and pathological ventricular repolarisation patterns from the 12-lead ECG and CMR-derived anatomies. We present a novel method that uses a flexible repolarisation parameterisation allowing both healthy and abnormal repolarisation patterns to be inferred from the T wave, as thoroughly evaluated against benchmark simulations with known ground truths. Building on previous approaches, this method incorporates as repolarisation parameters both the spatial distribution of APDs and a global AP shape, accounting for effects of AP triangulation in the T wave. Forward ECG simulation of many candidate repolarisation models was facilitated using a novel rapid repolarisation approximation, which combined a precomputed bivariate AP table indexed by APD tuples with tuned Gaussian smoothing to approximate reaction-diffusion. Finally, we demonstrated how errors propagate internally in the inference process, showing how an accurate activation inference supports a more accurate repolarisation inference.

## 4.1 Digital twinning of cardiac electrophysiology across healthy and diseased hearts

Previously, ventricular activation properties have been accurately inferred from the 12-lead ECG QRS complex using a simulation-based digital twinning approach, as evaluated against known ground-truth activation properties (Camps et al., 2021). The present study reinforces the accuracy of this approach in reconstructing ventricular activation times on six previously unseen ventricular activation sequences. We further applied optimisations such as precomputing activation times from the candidate early activation sites, as in other works (Pezzuto et al., 2021). Previous work had partially extended such an approach of simulation-based twinning to cardiac repolarisation, where the strength of imposed monotonic APD gradients (apicobasal, anteroposterior, interventricular, and transmural) were inferred from the 12-lead ECG T wave (Camps et al., 2025). The present study developed upon this work by, firstly, evaluating the inference method against known ground-truth repolarisation times and demonstrating its accuracy, and secondly using a flexible ventricular APD parameterisation which enabled inference of repolarisation times in cases of non-monotonic APD gradients like in cardiomyopathic heart disease.

To facilitate the simulation of many candidate pseudo-ECGs during the repolarisation inference process, previous work similarly used an AP table combined with smoothing to approximate reaction-diffusion (Camps et al., 2025). The present study expanded upon this by using a bivariate AP table indexed by ($APD_{90}$, $APD_{50}$) rather than just the $APD_{90}$, to better acknowledge the effects of shape during the plateau and repolarisation phases of the AP on T wave morphology. Moreover, the scale of Gaussian smoothing applied in the present study was tuned based on the myocardial conduction velocity inferred during activation inference, to account for the relationship between myocardial conductivity and diffusion of membrane potentials. These improvements in the approximation of ventricular repolarisation enabled inferred repolarisation times to accurately reflect the ground truth, despite the electrophysiology of benchmark simulations originating from a far more complex system of coupled reaction-diffusion ordinary and partial differential equations. Therefore, the novel fast surrogate approximations of cardiac ventricular repolarisation developed in the present study may be of interest more generally for rapid electrophysiology simulation.

Indeed, beyond ECG-based inference methods, alternative rapid methods to approximate reaction-diffusion have been developed, where in general these approximations involve a careful trade-off between runtime complexity and use case. For example, one such approximation combines an ionic model with a triple-Gaussian diffusion approximation to enable arrhythmia simulations to be run on coarse ventricular meshes (Barrios Espinosa et al., 2025).

Sequential inference of ventricular activation and repolarisation characteristics has previously been used for digital twinning in a recent previous study (Camps et al., 2025), however the effect of error in the inferred activation times upon the repolarisation inference process has not yet been demonstrated. The present study identified that error in the inferred activation sequence can propagate to error in the inference of the repolarisation sequence, which underscores the importance of efforts to improve the activation sequence inference (Pezzuto et al., 2021), and evaluate its accuracy against ground truth measurements. The propagation of activation inference error to repolarisation inference error was thought to occur via three key routes: (1) small differences between the amplitudes of the target QRS and the QRS associated with the inferred activation model subtly affected normalisation of the full ECG signal and thus the T wave discrepancy score; (2) error in the inferred myocardial conduction velocity directly translated into error in the strength of Gaussian smoothing of membrane potentials applied during repolarisation; and (3) the general instability associated with trying to introduce APD adjustments to compensate for the effect of erroneous activation times in repolarisation times. Error propagation via (2) may be mitigated by inferring myocardial conduction velocity to greater precision in future work, as the present study only inferred this to the nearest 10 $cm/s$. For the future application of digital twins constructed using simulation-based inference methods, future work should consider all sources of error and propagation in the inference process, of particular relevance in the context of arrhythmia simulations which may be sensitive to small errors in the inferred ventricular model (Corral-Acero et al., 2020; Niederer et al., 2020; Pathmanathan et al., 2024). Specifically, due to the close relationship between APDs and effective refractory periods, the inferred repolarisation field determines refractoriness gradients and thus arrhythmic risk in the cardiac digital twin (Martínez Díaz et al., 2024).

The testing of different regularisation techniques is a mainstay of ECG imaging research (Molero et al., 2024). Regularisation was not required in the present study for an accurate reconstruction of ventricular activation times, nor in previous works (Camps et al., 2021), yet the present study found that regularisation aided in the reconstruction of repolarisation times. This is explained by, firstly, that activation in some sense has a structural constraint imposed by the Dijkstra distance between ventricular sites; one site being active at time zero means that an adjacent site will be activated some short time afterwards. Such a rule was not present for repolarisation. Secondly, spatial heterogeneity in the time-course of the membrane potential during ventricular activation may be less pronounced than with repolarisation. The membrane potential of a single site during activation appears well approximated by a step function (due to the rapid AP upstroke) (Camps et al., 2021), whereas there may be significant flexibility in the membrane potential shape during repolarisation, which facilitates more complex ventricular repolarisation solutions that produce the same T wave, to be remediated by the regularisation term. Interestingly, the imposition of ventricular APD gradients in previous works may have prevented the need for an explicit regularisation term (Camps et al., 2025).

## 4.2 Limitations and future perspectives

The principal limitation of the present study is that repolarisation inference used an approximation of ventricular repolarisation for rapid pseudo-ECG simulation, where both the benchmark biventricular electrophysiology simulations and AP table used the ToR-ORd ventricular electrophysiology model. Although state-of-the-art and well validated against various *in vitro* AP recordings and clinical data (Tomek et al., 2019), future work should evaluate how well the AP table formed from this model can be applied to repolarisation inference in an *in vivo* population with clinically heterogeneous electrophysiology. To some extent, the present study accounted for this by using a novel global AP shape parameter which enabled the AP table to allow for AP shapes of variable triangular extent in the inference process. However, the additional clinical heterogeneity in resting membrane potentials, AP notch morphology and/or other miscellaneous repolarisation properties may need to be considered, particularly in the context of myocardial ischaemia (Gorgels et al., 2023). Moreover, if AP morphologies were spatially heterogeneous enough to affect the T wave, then the global AP shape parameter may need replacing with a local description of AP shapes, significantly increasing the complexity of repolarisation inference. Furthermore, in relation to digital twinning, although the repolarisation times may be accurately inferred, the estimation of the underlying ventricular ion channel distribution remains a challenge as different combinations of ion channel conductances can produce APs of similar shape and duration. Future work may consider rule-based methods to reconstruct an inferred APD field by varying ion channel conductances that are known to be spatially heterogeneous, thereby offering a framework for simulating pharmacologic interventions targeting specific ion channels (Loewe et al., 2014). Using ECGs at varying heart rates might further refine the possible distributions of ion channels, with careful consideration given to potential confounders such as the sympathetic response or episodes of transient ischaemia (Carmeliet, 1999; Coleman et al., 2024b; Doste and Bueno-Orovio, 2021).

Furthermore, in the proposed fast approximation of cardiac repolarisation, the spatial scale of smoothing applied to approximate diffusion was tied to the inferred myocardial conduction velocity, where the relationship between conduction velocity and diffusion of potentials was investigated in cable and slab monodomain simulations. It may be necessary to evaluate how well this relationship between conduction velocity and diffusion translates to *in vivo* hearts. Concomitantly, techniques to evaluate separately the approximations of reaction (AP table) and diffusion (Gaussian smoothing of potentials) directly against *ex vivo* or *in vivo* hearts may require development. Whether using the repolarisation inference method on simulated ground truth cases or clinical cases, we expect that improvements to the repolarisation approximation accuracy will further improve the accuracy of the repolarisation inference. Improvements are further motivated by the fact that the present fast repolarisation approximation applies Gaussian smoothing only after the ventricular activation phase is completed, which may limit attempts to simulate short QT where the activation and repolarisation phases may overlap to a significant extent (Weiss et al., 2005).

Although during the repolarisation inference no strict APD gradients were imposed, the type of APD gradients possible to reconstruct may be limited by the proposed mutation step (Supplementary Section 1.4). To ensure mutated repolarisation parameter sets were significantly altered at the mutation step (and thus result in a timely convergence), $APD_{90}$s were rescaled in ventricular regions defined by a radius of at least $1\ cm$. It is possible that APD gradients on a finer length scale may require mutations on a finer length scale, which could come at the expense of convergence times.

Related to this point is that the present study evaluated the repolarisation inference method on control and HCM cases, but the exact patterns of abnormal repolarisation clinically in HCM are as yet unknown, thus the HCM repolarisation benchmarks assumed ionic remodelling and prolonged repolarisation to be on a macroscopic length scale. Indeed, the purpose of the present study was to work towards a better spatial characterisation of abnormal repolarisation in cardiac disorders. Therefore, future work may analyse the relationship that the match between inferred model and target T waves has with the match between inferred and ground truth repolarisation times, to identify markers of an inaccurate inference in cases where the ground truth repolarisation times are unknown. In such cases, the mutation step could be adapted (e.g. use of a finer length scale for APD mutations or introduction of other miscellaneous APD gradient mutation types).

The present study parameterised ventricular activation with simultaneous activation of early activation sites, rather than allowing different sites to have different activation times, as in previous work (Camps et al., 2021). However, we do not interpret this inference method as necessarily inferring the exact positions of early activation sites, instead finding a set of early activation sites that broadly reconstruct the overall ventricular activation sequence. This can be visualised in Figure 3A (anterior view, left), where activation times determined by two early activation sites in the ground truth are well approximated by activation times arising from four inferred early activation sites. In practice, ventricular activation is determined by thousands of Purkinje-myocardium junctions (Camps et al., 2024; Grandits et al., 2025), but ventricle-scale activation can be well approximated by 6-10 early activation sites (Cardone-Noott et al., 2016; Durrer et al., 1970; Pezzuto et al., 2021).

Future work will consider the clinical translation of simulation-based repolarisation inference, for which there are numerous prerequisites which extend beyond the present study. These prerequisites include (1) effective clinical ECG preprocessing, including the removal of external beat-to-beat signal variability including noise, as well as effective handling of U waves if present, which could interfere with repolarisation inference particularly in HCM cases with prolonged QT interval (Johnson et al., 2011); (2) consideration of error propagation from the mesh reconstruction process from CMR into the repolarisation inference process; (3) consideration of personalised torso-derived lead positions and/or quantification of error propagation of lead position uncertainty (although the influence of such factors may be minor (Zappon et al., 2025)); (4) consideration of factors not included in the present study, such as the torso, ribs, organs, and ventricular motion; (5) the applicability of the proposed fast repolarisation surrogates to clinical repolarisation; and (6) the inclusion of myocardial fibrosis in the ventricular models, as informed by patient late gadolinium enhancement or other CMR modalities. Finally, the simulation-based inference method should be evaluated against repolarisation times obtained from invasive electroanatomic mapping, similar to that performed for the ventricular activation analogue (Giffard-Roisin et al., 2017; Pezzuto et al., 2021).

## 5. Conclusion

The present study developed a simulation-based inference framework that accurately reconstructed healthy and pathological ventricular repolarisation patterns from the 12-lead ECG, as evaluated on benchmark cardiac electrophysiology simulations. This was facilitated by a novel flexible repolarisation model parameterisation, and the novel use of a bivariate AP table and tuned Gaussian smoothing, which enabled rapid pseudo-ECG surrogate simulation of candidate repolarisation models during the inference process. The effect of error in the activation sequence inference on the repolarisation inference was further highlighted, underscoring the importance of advancements made in activation digital twinning (Camps et al., 2021; Grandits et al., 2025; Pezzuto et al., 2021; T. Grandits et al., 2024).

Altogether, this framework holds significant potential for future clinical translation, enabling personalised characterisation of ventricular repolarisation abnormalities from the standard clinical 12-lead ECG. Successful integration of such methods into clinical practice could enhance diagnosis, improve risk stratification, and provide targeted therapeutic intervention for a range of cardiac disorders where abnormal repolarisation is present.

## Software availability

The computational tools used to perform the inferences will be made publicly available upon publication at https://github.com/JamesAlecColeman/sim-based-inf. Benchmark cardiac electrophysiology simulations were performed using MonoAlg3D, which is an open-source solver (Sachetto Oliveira et al., 2018).

## CRediT authorship contribution statement

**James A Coleman:** Conceptualisation, Data curation, Formal analysis, Investigation, Methodology, Software, Validation, Visualisation, Writing – original draft, Writing – review & editing. **Julia Camps:** Conceptualisation, Methodology, Writing – review & editing. **Abdallah Hasaballa:** Methodology, Writing – review & editing. **Alfonso Bueno-Orovio:** Conceptualisation, Methodology, Funding acquisition, Project administration, Resources, Supervision, Writing – review & editing.

## Declaration of competing interest

The authors declare that they have no known competing financial interests or personal relationships that could have appeared to influence the work reported in this paper.

## Acknowledgements

This work was funded by the SMASH-HCM project under Innovate UK grant 10110728. Julia Camps acknowledges further support from a fellowship from "la Caixa" Foundation (LCF/BQ/PI25/12100029). The authors acknowledge the use of the University of Oxford Advanced Research Computing (ARC) facility (https://doi.org/10.5281/zenodo.22558).

For the purpose of open access, the authors have applied a Creative Commons Attribution (CC BY) public copyright licence to any Author Accepted Manuscript version arising from this submission.